\documentclass[%
 reprint,
 amsmath,amssymb,
 aps,
]{revtex4-1}

\usepackage{graphicx}
\usepackage{dcolumn}
\usepackage{bm}

\begin{document}

\preprint{APS/123-QED}

\title{\boldmath{$\alpha$}-\boldmath{$\beta$} and \boldmath{$\beta$}-\boldmath{$\gamma$} phase boundaries of solid oxygen\\ observed by adiabatic magnetocaloric effect}

\author{T. Nomura}
\email{t.nomura@issp.u-tokyo.ac.jp}
\author{Y. Kohama}%
\author{Y. H. Matsuda}%
\email{ymatsuda@issp.u-tokyo.ac.jp} 
\author{K. Kindo}%

\affiliation{%
Institute for Solid State Physics, University of Tokyo, Kashiwa, Chiba 277-8581, Japan 
}%

\author{T. C. Kobayashi}
\affiliation{%
 Department of Physics, Okayama University, Okayama 700-8530, Japan}%

\date{\today}

\begin{abstract}
The magnetic-field-temperature phase diagram of solid oxygen is investigated by the adiabatic magnetocaloric effect (MCE) measurement with pulsed magnetic fields.
Relatively large temperature decrease with hysteresis is observed at just below the $\beta$-$\gamma$ and $\alpha$-$\beta$ phase transition temperatures owing to the field-induced transitions.
The magnetic field dependences of these phase boundaries are obtained as $T_\mathrm{\beta\gamma}(H)=43.8-1.55\times10^{-3}H^2$ K and $T_\mathrm{\alpha\beta}(H)=23.9-0.73\times10^{-3}H^2$ K.
The magnetic Clausius-Clapeyron equation quantitatively explains the $H$ dependence of $T_\mathrm{\beta\gamma}$, meanwhile, does not $T_\mathrm{\alpha\beta}$.
The MCE curve at $T_\mathrm{\beta\gamma}$ is of typical first-order, while the curve at $T_\mathrm{\alpha\beta}$ seems to have both characteristics of first- and second-order transitions.
We discuss the order of the $\alpha$-$\beta$ phase transition and propose possible reasons for the unusual behavior.

\end{abstract}

\maketitle

\section{introduction}
Molecular oxygen, O$_2$ has a spin quantum number $S=1$ and behaves as a magnetic molecule.
In the condensed phases, magnetic interaction between O$_2$ molecules has an important role for the condensation energy in addition to van der Waals interaction \cite{81PRB_DeFotis,85JPSJ_Uyeda,02LTP_Freiman,04PR_Freiman}.
As decreasing temperature, antiferromagnetic (AFM) correlation develops and three phases which have different crystallographic and magnetic structures appear as $\gamma$, $\beta$, and $\alpha$.
The $\gamma$ phase (54.4--43.8 K, paramagnetic, cubic) is called as a plastic phase where molecules are rotating at certain lattice sites \cite{85JPSJ_Uyeda,73PRB_Samu}.
In the $\beta$ phase (43.8--23.9 K, short-range AFM, rhombohedral), molecular axis is ordered to one direction with large volume contraction.
The geometrical frustration due to the triangular lattice of the basal plane suppresses the long-range AFM ordering.
In the $\alpha$ phase (23.9-- K, long-range AFM, monoclinic), the frustration is lifted by the lattice deformation and the long-range order is realized \cite{87JCP_Jansen,87JCP_Jansen2,83PRB_Etters}.

The $\beta$-$\gamma$ phase transition is of first-order where larger entropy than fusion is released \cite{29JACS_Giauq,69LTP_Fage}.
In contrast, the order of the $\alpha$-$\beta$ phase transition is not clear despite many calorimetric studies \cite{04PR_Freiman,29JACS_Giauq,69LTP_Fage,96Cryo_Lip,98LTP_Lip}.
In recent studies, most authors are inclined toward the opinion of "first-order but close to second-order" \cite{01LTP_Pro,05LTP_Gomo}. 
However, the highest-resolved heat capacity measurement suggests that there is an intermediate phase between the $\alpha$ and $\beta$ phases \cite{96Cryo_Lip}, and the detail of the $\alpha$-$\beta$ phase transition is still controversial.

In the last 50 years, the pressure-temperature ($P$-$T$) phase diagram of solid oxygen has been extensively studied and four high-pressure phases ($\delta$, $\epsilon$, $\zeta$, $\eta$) were discovered \cite{04PR_Freiman,79CPL_Nicol,90JPC_Ruoff,04PRL_Santoro}.
On the other hand, the magnetic-field-temperature ($H$-$T$) phase diagram has not been studied until recently.
In 2014, $\theta$ phase of solid oxygen was discovered in ultrahigh magnetic field over 100 T using the single-turn coil (STC) technique \cite{14PRL_Nomura,15PRB_Nomura}.
This is the first observation of the field-induced phase transition of oxygen.
At that time, strangely, the field-induced $\alpha$-$\beta$ and $\beta$-$\gamma$ phase transitions were not observed \cite{15PRB_Nomura}, and the field dependences of these transition temperatures ($T_\mathrm{\alpha\beta}$, $T_\mathrm{\beta\gamma}$) have never been clarified experimentally \cite{04PR_Freiman}.
The $H$-$T$ phase diagram is the most fundamental information of the magnetic material for discussing thermodynamical property.
Therefore, the clarification of the $H$-$T$ phase diagram is an important issue for the oxygen-related science and technology.

Calorimetric measurement is straightforward to clarify the thermodynamical relation in the phase diagram.
Recently, adiabatic magnetocaloric effect (MCE) measurement is developed for the pulse magnetic field technique and applied for various kinds of materials  \cite{13RSI_Kihara,14PRB_Kihara,14PRB_Kohama}.
In the adiabatic MCE measurement, we measure the $H$ dependence of $T$ which changes to conserve the total entropy.
In other words, each MCE curve corresponds to the isentropic curve in the $H$-$T$ phase diagram.
A contour plot of entropy enables the quantitative discussion on the entropy relation between the phases.

In this study, we conducted the adiabatic MCE measurement of condensed oxygen up to 56 T using pulse magnetic fields.
From the obtained isentropes, the magnetic field dependences of $T_\mathrm{\beta\gamma}$ and $T_\mathrm{\alpha\beta}$ were revealed for the first time.
We comment on the reason why these phase boundaries had not been observed in the previous high-field measurements \cite{85JPSJ_Uyeda,14PRL_Nomura,15PRB_Nomura}.
We also comment on the long lasting problem, the order of the $\alpha$-$\beta$ phase transition.

\section{experimental setup}
The concept of the adiabatic MCE measurement is described in Ref. \cite{13RSI_Kihara}.
The MCE data were obtained in the pulsed magnetic fields with the duration time of 36 ms.
Two types of resistance thermometers, Cernox bare chip (CX-1050 or CX-1030, Lake Shore Cryotronics) and RuO$_2$ film (EZ-13, Tanaka Kikinzoku Kogyo K. K.) were employed.
The sizes of the Cernox and the RuO$_2$ film were $1\times0.8\times0.2\ \mathrm{mm}^3$ and $2\times1.5\times0.1\ \mathrm{mm}^3$, respectively.
RuO$_2$ was employed only for the low temperature region below 10 K where the reliability becomes better than Cernox.
Resistance was measured by the standard ac four-probe method using numerical lock-in technique at the frequency of 100 kHz.

\begin{figure}[bt]
\centering
\includegraphics[width=7cm]{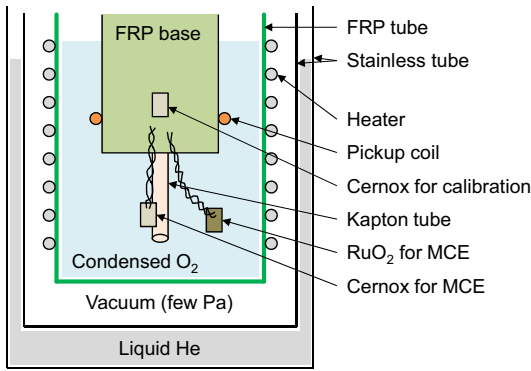}
\caption{\label{fig:MCE_setting}
Schematic setting of the MCE measurement for condensed oxygen.
}
\end{figure}

Schematic setup near the thermometer is shown in Fig. \ref{fig:MCE_setting}.
Solid oxygen was condensed from high-purity O$_2$ gas (99.999\%) at the bottom of the tube made of fiber-reinforced plastic (FRP).
The inner diameter of the tube was 10 mm.
Two thermometers were directly buried inside the condensed oxygen for ideal thermal contact.
The thermometers located at 10 mm below the FRP base to reduce the heat transfer from the construction parts and to realize the adiabatic condition.
The Cernox was fixed on the Kapton tube (diameter 1 mm, thickness 0.06 mm) by vanish to suppress mechanical vibration.
The RuO$_2$ film was freely hanged in the condensed oxygen.
The thermometers are rigidly fixed inside solid oxygen at below 43.8 K.

The temperature dependence of the ac resistance is calibrated for each setup using the calibrated Cernox at zero field.
The effect of magnetoresistance at each temperature is estimated from the same measurement for solid argon where the intrinsic temperature change ($\Delta T$) is negligible.
The artificial $\Delta T$ caused by magnetoresistance of the thermometer is subtracted as background.
In this analytical process, non-linearity and angle dependence of the magnetoresistance could cause large error of $\Delta T$ \cite{99RSI_Brandt}.
The relative error of $\Delta T$ at 50 T is estimated as $\pm0.2$ K.
The absolute error of $T$ is estimated as $\pm0.5$ K.
The absolute error mainly originates from the inhomogeneity of temperature during the calibration of the thermometers.
Near the phase boundaries of solid oxygen, the absolute error of $T$ can be recalibrated by $T_\mathrm{\beta\gamma}=43.8$ K and $T_\mathrm{\alpha\beta}=23.9$ K where heat capacity diverges.
These values are recommended for the thermometric fixed points \cite{98LTP_Lip}.

\section{result}
\begin{figure}[bth]
\centering
\includegraphics[width=8.6cm]{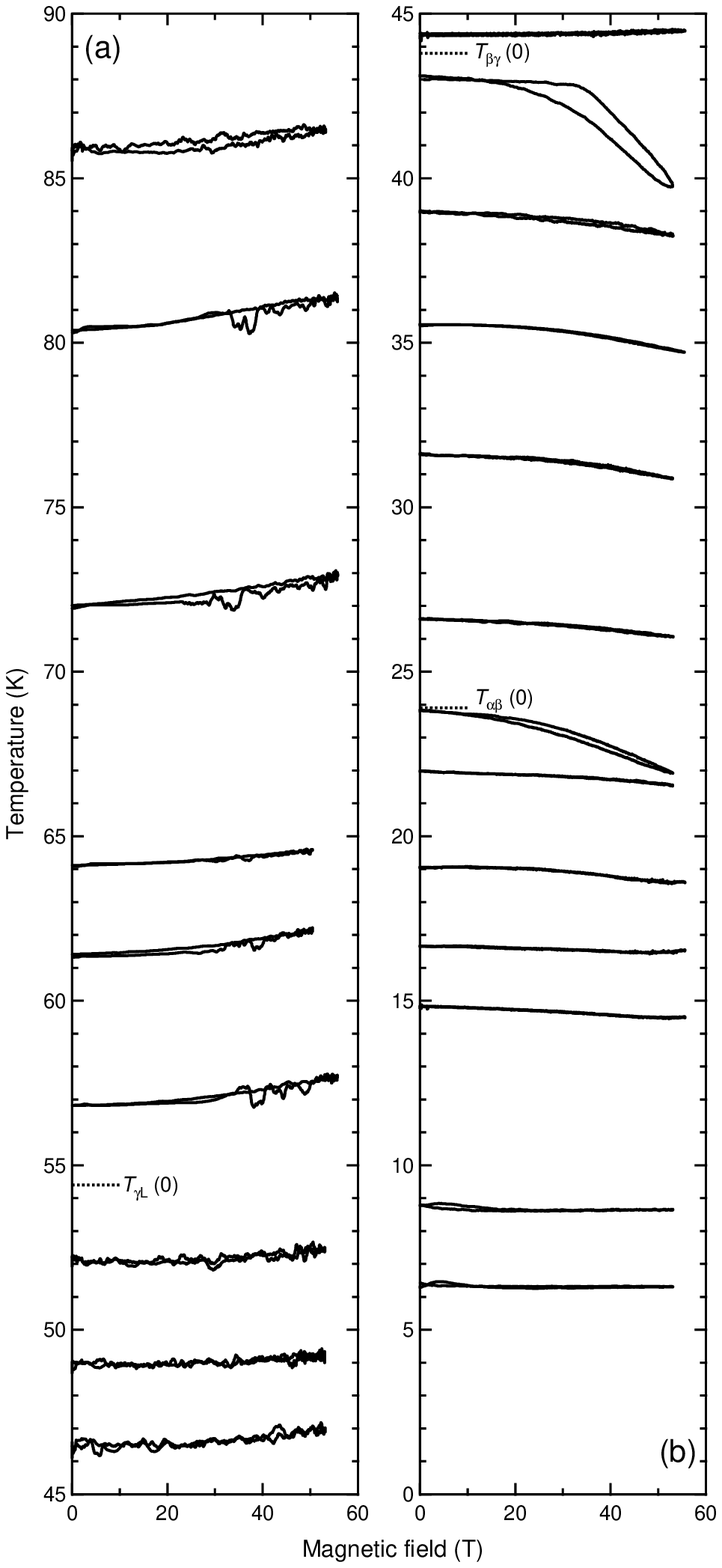}
\caption{\label{fig:MCE_AllT}
MCE curves of condensed oxygen in (a) 90-45 K and in (b) 45-0 K.
Phase boundaries at zero field ($T_\mathrm{\gamma\mathrm{L}}(0)$, $T_\mathrm{\beta\gamma}(0)$, $T_\mathrm{\alpha\beta}(0)$) are shown by dotted lines.}
\end{figure}

Summarized results of the MCE measurement of condensed oxygen are shown in Fig. \ref{fig:MCE_AllT} for (a) higher and (b) lower temperature regions.
The $\gamma$-liquid, $\beta$-$\gamma$, and $\alpha$-$\beta$ phase boundaries at zero field ($T_\mathrm{\gamma\mathrm{L}}(0)$, $T_\mathrm{\beta\gamma}(0)$, $T_\mathrm{\alpha\beta}(0)$) are shown by dotted lines.
Larger noise level in high temperature region is due to the mechanical vibration and relatively lower sensitivity of the thermometer.
The effect of vibration is suppressed at low temperature by the solidification of oxygen.

\begin{figure}[bth]
\centering
\includegraphics[width=7.8cm]{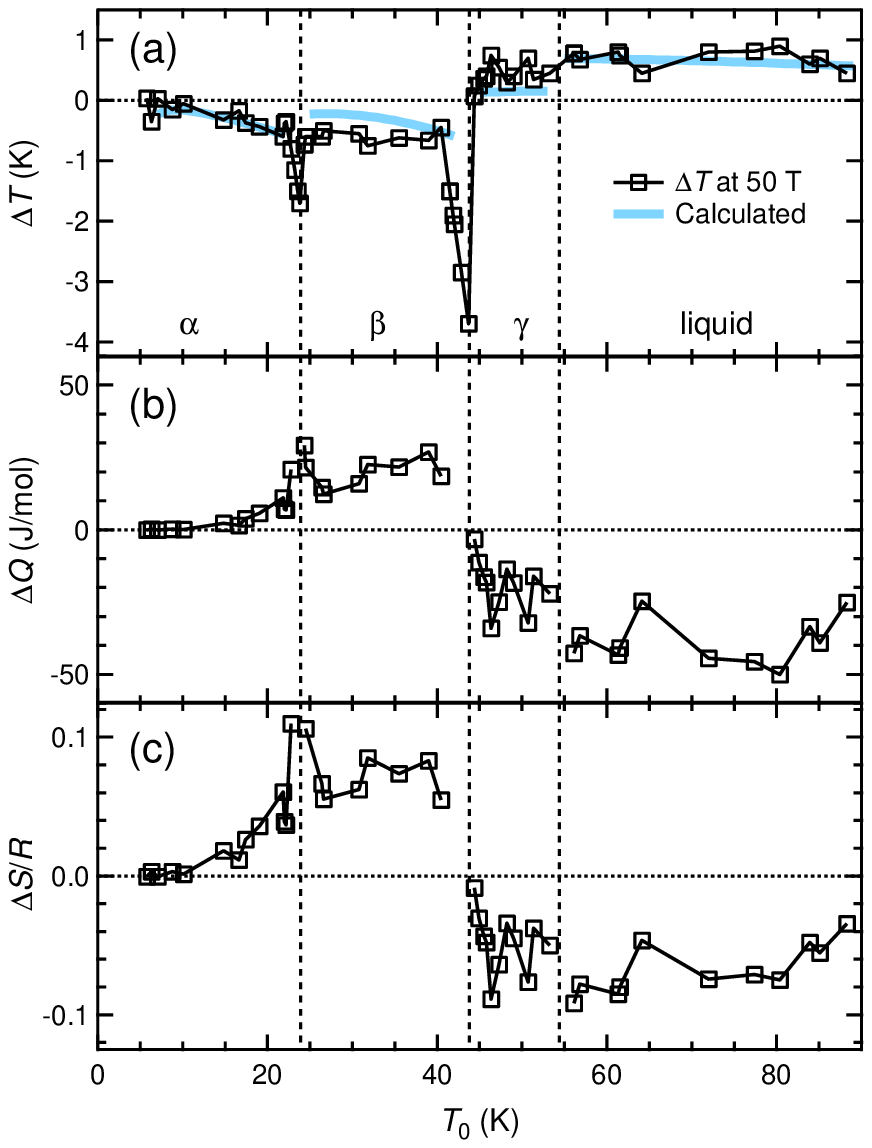}
\caption{\label{fig:MCE_dT_dQ_dS}
(a) Temperature changes at 50 T as a function of $T_0$.
The blue curve shows the estimated value by Eq. (\ref{eq:dMCE_int}).
Corresponding amounts of (b) heat, and (c) entropy.
Entropy is normalized by the gas constant, $R=8.31$ JK$^{-1}$mol$^{-1}$.
}
\end{figure}

$\Delta T$ for each initial temperature ($T_0$) is summarized in Fig. \ref{fig:MCE_dT_dQ_dS} (a).
Corresponding amounts of heat ($\Delta Q$) and entropy change ($\Delta S$) are shown in Figs. \ref{fig:MCE_dT_dQ_dS} (b) and (c), respectively.
The definitions are as follows.
\begin{equation}
\Delta Q=-c_0 \Delta T,
\end{equation}
\begin{equation}
\Delta S=\Delta Q/T_0 ,
\end{equation}
where $c_0$ is heat capacity at $T_0$ and zero field reported by Fagerstroem and Hollis Hallett \cite{69LTP_Fage}.
Data plots near the phase boundaries are removed since heat capacity greatly depends on $T$ and $H$.
$\Delta T$ at 50 T is less than 1 K in most $T_0$ region.
Only near the $\beta$-$\gamma$ and $\alpha$-$\beta$ phase boundaries, larger temperature decrease is observed with hysteresis.
This indicates that the $\beta$-$\gamma$ and $\alpha$-$\beta$ phase transitions are induced by magnetic fields.

\begin{figure}[bth]
\centering
\includegraphics[width=7cm]{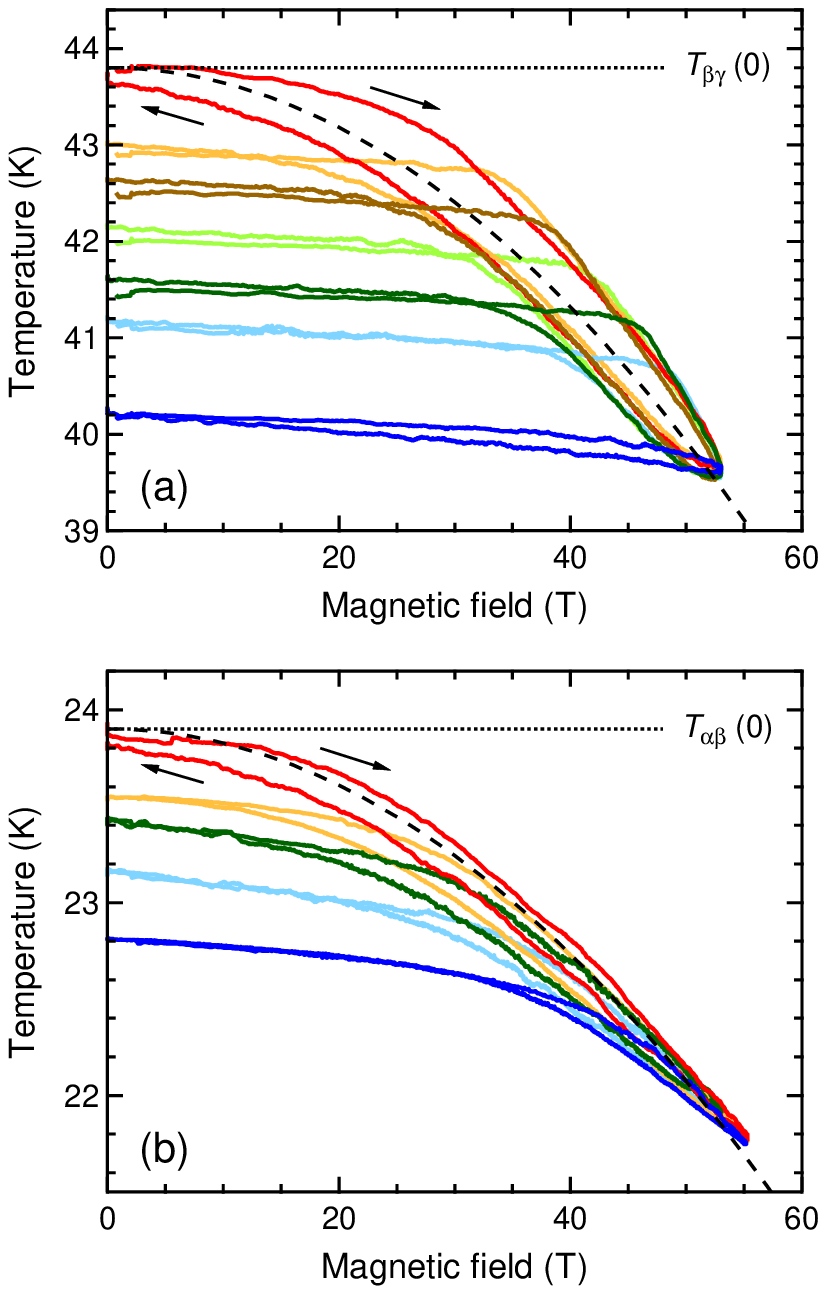}
\caption{\label{fig:MCE_bg_ab}
Enlarged MCE results at around (a) $T_\mathrm{\beta\gamma}$ and (b) $T_\mathrm{\alpha\beta}$.
Black dashed curves are the quadratic function shown for the phase boundaries.
}
\end{figure}

The enlarged MCE curves with different $T_0$ are shown in Figs. \ref{fig:MCE_bg_ab} for the (a) $\beta$-$\gamma$ and (b) $\alpha$-$\beta$ boundaries.
Even if $T_0$ is slightly changed, all MCE curves reach to the same temperature at the top of the field and go back to $T_0$ with hysteresis.
By connecting the center of hysteresis with quadratic function, the magnetic field dependence of the phase boundaries are obtained as the dashed curves.

Noteworthy, the MCE curves at around the $\alpha$-$\beta$ phase boundary depend on the thermal history.
When the $\alpha$ phase is prepared with keeping the temperature above 20 K, the MCE curves become different from Fig. \ref{fig:MCE_bg_ab} (b) although the obtained phase boundary is similar (Supplemental Material S-I \cite{supple}).
Lipinski \textit{et al.} also pointed out that the sample of the $\alpha$ phase has to be prepared at below 20 K to obtain reproducible data of heat capacity \cite{96Cryo_Lip,98LTP_Lip}.
In this paper, all data near the $\alpha$-$\beta$ phase boundary were collected with the sample which experienced the temperature below 20 K.

In this study, there was no indication of the field-induced $\gamma$-liquid phase transition near $T_\mathrm{\gamma\mathrm{L}}(0)$.
This would be because $T_\mathrm{\gamma\mathrm{L}}$ is almost independent on $H$ \cite{15PRB_Nomura}.
Since the magnetic susceptibilities of the $\gamma$ and liquid oxygen are only slightly different \cite{85JPSJ_Uyeda}, external magnetic field can not be a driving force of this phase transition.
$T$ sweep at fixed $H$ would be necessary to study the phase boundary.

\section{discussion}
At first, we discuss the common results of the MCE where no phase transition occurs.
The magnetocaloric relation \cite{69PR_Schelleng,69JAP_Butterwoth}
\begin{equation}
(\frac{\partial T}{\partial H})_S=-\frac{TH}{c_H}(\frac{\partial \chi}{\partial T})_H,
\label{eq:dMCE}
\end{equation}
where $c_H$ is the specific heat at constant magnetic field, manifests that the sign of $\partial \chi/\partial T$ determines the sign of MCE.
$\partial \chi/\partial T$ is negative for the liquid and $\gamma$, and positive for the $\beta$ and $\alpha$ phases \cite{81PRB_DeFotis}.
Therefore, $\Delta T$ in Fig. \ref{fig:MCE_dT_dQ_dS} (a) qualitatively agrees with Eq. (\ref{eq:dMCE}).

To a first approximation, $\Delta T$ is obtained by fixing $c_H=c_0$, $T=T_0$, and $({\partial \chi}/{\partial T})_{H=0}$ as
\begin{equation}
\Delta T=-\frac{T_0({\partial \chi}/{\partial T})_{H=0}}{2 c_0}H^2.
\label{eq:dMCE_int}
\end{equation}
The expected $\Delta T$ at 50 T is shown by the blue curve in Fig. \ref{fig:MCE_dT_dQ_dS} (a) (details are given in Supplemental Material S-II \cite{supple}).
It quantitatively agrees for all phases, but is not perfect for the $\gamma$ and $\beta$ phases.
That means the approximation fails for these phases, at most by the factor of two.
Most probably, it is due to the inaccuracy of $\partial \chi/\partial T$ under the external magnetic field.

In the following sections, we quantitatively discuss the $\beta$-$\gamma$ and $\alpha$-$\beta$ phase boundaries.
For estimating the entropy at zero field, the specific heat data reported by Fagerstroem and Hollis Hallett \cite{69LTP_Fage} are employed (Supplemental Material S-III \cite{supple}).

\subsection{\boldmath{$\beta$-$\gamma$} phase boundary}
Here, we discuss the results near the $\beta$-$\gamma$ phase boundary in Fig. \ref{fig:MCE_bg_ab} (a). 
Even when $T_0$ changes, all MCE curves follow the same boundary (dashed curve) and go back to $T_0$ with hysteresis.
This behavior is due to the phase equilibrium between the $\beta$ and $\gamma$.
When the magnetic field reaches to the first-order $\beta$-$\gamma$ phase boundary, the $\beta$ starts to transform to the $\gamma$.
Here, the transition occurs only partially since the total entropy is conserved in the adiabatic condition.
When the fraction of the $\gamma$ phase increases, temperature has to decrease to compensate the entropy difference $\Delta S_\mathrm{\beta\gamma}=2.04R$ \cite{69LTP_Fage}. 
Thus, the total entropy is conserved by balancing the fraction and temperature along the $\beta$-$\gamma$ phase boundary.
Because of this balance, all MCE curves reach to the same point ($H=53$ T, $T=39.6$ K) regardless of the different $T_0$.
In other words, the magnetic field of 53 T is not enough to overcome the entropy barrier $\Delta S_\mathrm{\beta\gamma}$ and to transform the entire $\beta$ phase into the $\gamma$ phase.

The fraction of the $\gamma$ phase at 53 T can be estimated by the equation of entropy.
We write the entropy as functions of $H$ and $T$ for the $\beta$ and $\gamma$ phases as $S_\beta (H, T)$ and $S_\gamma (H, T)$, respectively.
The initial entropy is equal to the average of them as
\begin{equation}
S_{\beta}(\mathrm{0\ T}, T_0)=c_{\beta}S_{\beta}(H, T) +c_{\gamma}S_{\gamma}(H, T),
\label{eq:b_eq_betagamma}
\end{equation}
where $c_\beta$ and $c_\gamma$ are the fractions of the $\beta$ and $\gamma$ phases, respectively.
Here, the contribution of mixing entropy is neglected.
By introducing the entropy difference between the $\beta$ and $\gamma$ phases $\Delta S_\mathrm{\beta\gamma} (H, T)$ and using the relation of $c_\beta = 1-c_\gamma$,
\begin{equation}
c_{\gamma}=\frac{S_\beta(\mathrm{0\ T}, T_0)-S_\beta(H, T)}{\Delta S_\mathrm{\beta\gamma}(H, T)}.
\label{eq:b_eq_d_betagamma1}
\end{equation}
The entropy difference between the $\beta$ and $\gamma$ phases at zero field is reported as 2.04$R$ \cite{69LTP_Fage}.
Figure \ref{fig:MCE_dT_dQ_dS} (c) shows that $\Delta S_\mathrm{\beta\gamma}(H, T)$ decreases approximately by 0.1$R$ at 50 T.
Therefore, $\Delta S_\mathrm{\beta\gamma} (\mathrm{53\ T}, \mathrm{39.6\ K}) = 2.04R - 0.1R = 1.94R$.
By using the isentropic relation (blue curve in Fig. \ref{fig:MCE_bg_ab}(a)), $S_\beta(\mathrm{53\ T}, \mathrm{39.6\ K})=S_\beta(\mathrm{0\ T}, \mathrm{40.2\ K})=3.88R$ \cite{supple}. 
If we consider the case of $T_0 = 43.0$ K, the initial entropy is $S_\beta(\mathrm{0\ T}, \mathrm{43.0\ K})=4.22R$ \cite{supple}.
$c_\gamma$ at the top of the field is obtained as $c_\gamma = 0.18$.
Even at 53 T, the $\beta$-$\gamma$ phase transformation occurs only partially under the adiabatic condition.
If $T_\mathrm{\beta\gamma}$ continues to decrease with the dashed curve, 100$\pm 10$ T is necessary for $c_\gamma = 1$.

This is the reason why the field-induced $\beta$-$\gamma$ phase transition had never been observed in the early pulsed-field experiments \cite{85JPSJ_Uyeda,14PRL_Nomura,15PRB_Nomura}.
They were conducted at the adiabatic condition because of the short duration of the field.
In the optical and magnetization measurements, averaged results from the coexisting $\beta$ and $\gamma$ phases are obtained.
If $c_\gamma$ gradually increases, it is difficult to detect the phase transition and separate the contributions from coexisting phases.
We emphasize that this phase coexistence is only the case for adiabatic condition.
If the system is isothermal, thermal energy is provided by the heat bath and the phase transition finishes at a certain field.

Next, we compare the obtained phase boundary with the expected one derived from the thermodynamical relation.
The slope of the phase boundary is described by the magnetic Clausius-Clapeyron equation
\begin{equation}
dT_\mathrm{c}/dH_\mathrm{c} =-\Delta M/\Delta S=-\Delta \chi H/\Delta S.
\label{eq:cc_eq}
\end{equation}
Here, $\Delta \chi$ is the differences of magnetic susceptibility between two phases.
If the ratio $\lambda=\Delta \chi/2 \Delta S$ is independent on $H$, an integrated form is written as
\begin{equation}
T_\mathrm{c}(H) = T_\mathrm{c}(0)-\lambda H^2.
\label{eq:cc_eq_INT}
\end{equation}
The $\beta$-$\gamma$ phase boundary is well fitted by this formula as
\begin{equation}
T_\mathrm{\beta\gamma}(H)=43.8-1.55\times10^{-3}H^2.
\label{eq:bg_boundary_fitted}
\end{equation}
By using the reported values of $\Delta \chi_\mathrm{\beta\gamma}=51.2\times10^{-3}$ JT$^{-2}$mol$^{-1}$ \cite{81PRB_DeFotis} and $\Delta S_\mathrm{\beta\gamma}=16.9$ JK$^{-1}$mol$^{-1}$ \cite{69LTP_Fage}, $\lambda$ at zero field is estimated as $\lambda=1.51\times 10^{-3}$ KT$^{-2}$.
This value is in good agreement with the experimental one.

The correspondence between the experiment and Eq. (\ref{eq:cc_eq_INT}) means that $\lambda$ is independent on $H$.
However, Fig. \ref{fig:MCE_dT_dQ_dS}(c) shows that $\Delta S_\mathrm{\beta\gamma}$ decreases as $H$ increases.
For the compensation, it is indicated that $\Delta \chi_\mathrm{\beta\gamma}$ also decreases in magnetic fields.
This is explained by the Brillouin-like magnetization curve of the paramagnetic $\gamma$ phase.
When the magnetization curve of the $\gamma$ phase shows a trend of saturation in high fields, $\Delta \chi_\mathrm{\beta\gamma}$ decreases.
For the case of the $\beta$-$\gamma$ phase boundary, these contributions compensate and $\lambda$ fortuitously stays constant.

\subsection{\boldmath{$\alpha$-$\beta$} phase boundary}
Compared with the $\beta$-$\gamma$ phase boundary, the MCE curves near the $\alpha$-$\beta$ phase boundary are difficult to interpret since they have both characteristics of first- and second-order transitions.
As a characteristic of the first-order, all MCE curves reach to the same point ($H=56$ T, $T=21.8$ K) with slight hysteresis, indicating that the $\alpha$ and $\beta$ phases are coexisting.
However, one unique phase boundary can not obtained by connecting the center of hysteresis.
This is because the temperature decrease with hysteresis starts before it reaches the $\alpha$-$\beta$ phase boundary.
This behavior resembles with the second-order (continuous) transition which shows continuous temperature change prior to the phase transition.
In Fig. \ref{fig:MCE_bg_ab}(b), we propose a feasible phase boundary by the dashed curve as
\begin{equation}
T_\mathrm{\alpha\beta}(H)=23.9-0.73\times10^{-3}H^2.
\label{eq:ab_boundary_fitted}
\end{equation}

We compare the obtained $\alpha$-$\beta$ phase boundary with the predicted one by Jansen and Avoird \cite{87JCP_Jansen2}.
Their prediction was based on Eq. (\ref{eq:cc_eq_INT}), and the parameters were employed as $\Delta \chi_\mathrm{\alpha\beta}=14.1\times10^{-3}$ JT$^{-2}$mol$^{-1}$ \cite{81PRB_DeFotis} and $\Delta S_\mathrm{\alpha\beta}=3.85$ JK$^{-1}$mol$^{-1}$ \cite{66SJPC_Orlova}.
$\lambda$ at zero field was predicted as $\lambda=1.8\times 10^{-3}$ KT$^{-2}$.
This value is more than twice the value obtained in this study.

The discrepancy is considered to be due to the inaccuracy of $\Delta \chi_\mathrm{\alpha\beta}$ and $\Delta S_\mathrm{\alpha\beta}$.
The Clausius-Clapeyron equation can be applicable for any points in the phase diagram.
However, $\Delta \chi$ and $\Delta S$ have to be estimated for the infinitesimal $\Delta T$.
This estimation is difficult near the $\alpha$-$\beta$ phase transition since $\chi$ and $S$ show tendency of divergence.
Actually, reported $\Delta S_\mathrm{\alpha\beta}$ differs each other by around 20\% even if we exclude the reports of "no latent heat" \cite{04PR_Freiman}.
Therefore, the employed values of $\Delta \chi_\mathrm{\alpha\beta}$ and $\Delta S_\mathrm{\alpha\beta}$ could be inappropriate for the estimation.
The ratio obtained in this study ($\lambda=\Delta \chi/2 \Delta S=0.73\times 10^{-3}$ KT$^{-2}$) is considered to be more reliable.

Next, we discuss the long-lasting problem of solid oxygen, the order of the $\alpha$-$\beta$ phase transition.
The history of the controversy is well summarized in Ref. \cite{96Cryo_Lip}.
After 1990, most researches insist first-order by the measurements of heat capacity \cite{96Cryo_Lip,98LTP_Lip}, x-ray diffraction \cite{01LTP_Pro}, optical spectroscopy \cite{00LTP_Minenko,01PRB_Medv}.
Our results of the adiabatic MCE measurements, which indicate two-phase coexistence with hysteresis, agree with these researches.
However, the reason of why the MCE curves also show the continuous-transition-like behavior is not clear.
In the following discussion, three possible reasons for this behavior are proposed.

The first possible reason is inhomogeneous stress in the sample.
The sample used in this study is polycrystalline.
The local stress at the grain boundaries could slightly change the transition field.
Especially, the $\alpha$-$\beta$ phase transition is considered to be martensitic \cite{85JPSJ_Uyeda,67PR_Barrett0,67PR_Barrett}, which implies sensitive to stress.
However, we confirmed that the MCE curves in Fig. \ref{fig:MCE_bg_ab} (b) were reproduced in four different samples with different settings.
If it originates from local stress, it should depend on the sample quality and show different behaviors for each sample.
Therefore, the effect of local stress does not seem to be the dominant reason.

The second one is the magnetic anisotropy of the AFM ordered $\alpha$ phase \cite{87JCP_Jansen2}.
Since the magnetic susceptibility of the $\alpha$ phase is anisotropic, the transition field could vary depending on the orientation of each domain.
The diffuse $\alpha$-$\beta$ phase boundary for the polycrystalline sample was proposed in Ref. \cite{87JCP_Jansen2} even though it does not coincide with Eq. (\ref{eq:ab_boundary_fitted}).
In this sense, the $\alpha$-$\beta$ phase transition should take place in broad area of the $H$-$T$ phase diagram for polycrystal.

The third one is the intermediate phase between the $\alpha$ and $\beta$ phases.
The high-resolution measurement of heat capacity indicated that the peak of the $\alpha$-$\beta$ phase transition is composed of two sharper peaks with the separation of 0.02 K \cite{96Cryo_Lip,98LTP_Lip}.
The authors argued that the double peak is due to the intermediate phase.
Interestingly, the helical-ordered intermediate phase is theoretically predicted by Slusarev \textit{et al.} \cite{Cpeaks80,Cpeaks81} and discussed by Gaididei and Loktev \cite{ab_theory81}.
However, the existence of the intermediate phase has not been confirmed in other measurements because of its tiny temperature range.
If the intermediate phase exists, the behavior of the MCE curve is not clear; it depends on the orders of $\alpha$-intermediate and intermediate-$\beta$ transitions.
Moreover, the equilibrium is not guaranteed for the pulsed field measurement.
In any cases, the existence of the intermediate phase could affect the MCE curve in an unusual way.

At this stage, we cannot conclude which is the main factor to explain the MCE curve of the $\alpha$-$\beta$ phase transition.
For further discussions, single crystal of the solid oxygen $\alpha$ phase is necessary.
For single crystal, the effect of strain would be suppressed and the transition field can be discussed for each axis.
For the third reason, further investigation is difficult in the pulse field because the precise control of $T$ and $H$ is necessary.

\section{conclusion}
The adiabatic MCE measurement was conducted for liquid and solid oxygen up to 56 T.
$\Delta T$ was qualitatively discussed for each phase in terms of the magnetocaloric relation.
Relatively large temperature decrease with hysteresis was observed at just below $T_\mathrm{\beta\gamma}(0)$ and $T_\mathrm{\alpha\beta}(0)$, suggesting the $\beta$-$\gamma$ and $\alpha$-$\beta$ phase transitions occurred.
In the adiabatic condition, the entropy of the system is conserved.
When the high-entropy phase is induced by magnetic fields, temperature decreases for compensating the entropy difference at the first-order transition.
At that time, the fraction of the high-field phase gradually increases along the phase boundary with decreasing $T$.
This is the reason for that the field-induced $\beta$-$\gamma$ and $\alpha$-$\beta$ phase transitions had not been observed in pulsed magnetic fields.
The obtained $\beta$-$\gamma$ phase boundary was quantitatively explained by the magnetic Clausius-Clapeyron equation, while the $\alpha$-$\beta$ phase boundary was not.
The discrepancy of the $\alpha$-$\beta$ phase boundary was attributed to the difficulty of estimating $\Delta \chi_\mathrm{\alpha\beta}$ and $\Delta S_\mathrm{\alpha\beta}$.

The MCE curve at the $\beta$-$\gamma$ phase boundary is of typical first-order transition.
On the other hand, the MCE curve at the $\alpha$-$\beta$ phase boundary has both characteristics of first- and second-order transitions.
We argued this could be due to the polycrystalline sample or the intermediate phase between the $\alpha$ and $\beta$ phases.
In any cases, our results agree with the previous studies which suggest that the $\alpha$-$\beta$ phase transition is of first-order.

\section*{acknowledgments}
TN was supported by Japan Society for the Promotion of Science through the Program for Leading Graduate Schools (MERIT) and a Grant-in-Aid for JSPS Fellows.
This work was partly supported by JSPS KAKENHI, Grant-in-Aid for Scientific Research (B) (16H04009).


\end{document}